\documentclass[conference, compsoc, twoside]{IEEEtran}
\usepackage{fancyhdr}
\usepackage{graphicx}
\usepackage{multirow}

\ifCLASSOPTIONcompsoc
  \usepackage[nocompress]{cite}
\else
  \usepackage{cite}
\fi
\ifCLASSINFOpdf
\else
\fi
\begin{document}

%

\title{Multi-task Neural Networks for Personalized Pain Recognition \\from Physiological Signals}

\author{\IEEEauthorblockN{Daniel Lopez-Martinez}
\IEEEauthorblockA{Harvard-MIT Division of Health Sciences and Technology\\
Massachusetts Institute of Technology\\
Cambridge, USA \\
Email: dlmocdm@mit.edu}

\and
\IEEEauthorblockN{Rosalind Picard}
\IEEEauthorblockA{MIT Media Lab\\
Massachusetts Institute of Technology \\
Cambridge, USA\\
Email: picard@media.mit.edu}}


%


\maketitle
\thispagestyle{fancy}
\begin{abstract}
Pain is a complex and subjective experience that poses a number of measurement challenges. While self-report by the patient is viewed as the gold standard of pain assessment, this approach fails when patients cannot verbally communicate pain intensity or lack normal mental abilities. Here, we present a pain intensity measurement method based on physiological signals. Specifically, we implement a multi-task learning approach based on neural networks that accounts for individual differences in pain responses while still leveraging data from across the  population. We test our method in a dataset containing multi-modal physiological responses to nociceptive pain.
\end{abstract}


%
\IEEEpeerreviewmaketitle

\section{Introduction}

Pain is an unpleasant sensory and emotional experience associated with actual or potential tissue damage with sensory, emotional, cognitive and social components \cite{DeCWilliams2016}. While self-rating pain assessment tools such as the numerical rating scale (NRS) or the visual analogue scale (VAS) are the most common measures of pain intensity used by clinicians and researchers \cite{Younger2009}, these methods only work when the subject is sufficiently alert and cooperative. Therefore, automatic recognition of pain is of increased interest in situations in which the severity of pain cannot be communicated, such as when the subject is either drowsy or unconscious, or in special patient populations with verbal and/or cognitive impairments.

While recent advances in automatic pain recognition have mainly focused on pain intensity prediction from behavioral cues such as facial expressions (for an example, see \cite{LopezMartinez2017c}), physiological signals can also be used for pain recognition \cite{Chu2017,Kachele2016}. Specifically, pain has been shown to interact with the autonomic nervous system (ANS) \cite{benaroch2001} and hence to lead to changes in skin conductance (SC) and  heart rate (HR)  \cite{Chu2017,Walter2014a}. While most previous automatic approaches for pain recognition have also used electromyograms (EMG) \cite{Walter2014a,Werner2015a,Kachele2016} or electroencephalograms (EEG) \cite{Zhang2012}, the attachment of electrodes to the face and scalp is not suitable for practical applications. 
Skin conductance and heart rate, on the other hand, can be conveniently measured  from wearable sensors placed on the wrist.

In this paper, we focus on the recognition of nociceptive pain using skin conductance and heart rate data. Since pain has been shown to elicit different physiological responses in different people (e.g. due to gender or age differences)  \cite{Ledowski2011,Tousignant-laflamme2005,Tousignant-Laflamme2006}, we adopt a multi-task learning (MTL) approach with person-specific outputs \cite{Jaques15} to account for these individual differences in pain responses. Specifically, we implement a MTL neural network architecture with hard parameter sharing (see Figure \ref{fig:nnarchitecture}) \cite{Ruder2017}. Our model is therefore able to account for individual differences while still learning from the data of other subjects through shared layers in the neural network.

The main contributions of this work are: (1) use of  neural networks, (2) use of a multi-task learning approach, and (3) use of multi-modal data (skin conductance and heart rate) for personalized nociceptive pain recognition in healthy subjects.



\section{Related work}


While self-report by the patient is viewed as the gold standard of pain assessment, this approach fails when patients cannot verbally communicate pain intensity. Therefore, developing an objective, sensitive, specific, continuous, and online method to monitor pain has recently become a focus of work. To this end, many studies have explored different physiological indicators of pain. Variations in physiological parameters such as skin conductance \cite{Storm2008}, heart rate \cite{Jess2016}, blood pressure \cite{Sacco2013}, pulse oximetry \cite{Hamunen2012} and brain hemodynamics \cite{Aasted2016} have been characterized in different perioperative settings and in healthy human experimental pain models, in which the assay by which pain is assessed involves a nociceptive pain stimulus that can be electrical, thermal (heat, cold), mechanical (blunt, punctate pressure), or chemical (intranasal CO$_2$, nociceptive substances) \cite{Lotsch2014}. However, the majority of these studies examined the relationships between pain  and a single parameter. Given that these physiological signals represent different systems, the parameters derived from them are complementary, rather than redundant \cite{Treister2012}. 

To date, there are only a few works that have applied machine learning to the problem of automatic pain detection using multi-modal data. 
For example, Chu et al. \cite{Chu2014} used blood volume pulse (BVP), ECG and SC features in a linear discriminant analysis (LDA) classifier  to discriminate between seven physiological states: five levels of electrical pain, 
pre-stimulaton and post-stimulation. Subsequently, $k$-nearest neighbor and support vector machine (SVM) classifiers were also applied to the same multi-modal data \cite{Chu2017}. 
Walter et al. \cite{Walter2014a} used EMG, ECG, SC and EEG features to classify four levels of heat pain against baseline (no pain) using SVMs, in the BioVid Heat Pain dataset \cite{Walter2013}. Gruss et al. \cite{Werner2015a} also applied SVMs to the same dataset, but only used EMG, ECG and SC features. These features have also been used in combination with behavioral features derived from video \cite{Werner2014a,Kachele2015,Iliadis2015}. Recently, Kächele et al. \cite{Kachele2016} proposed a method for personalized prediction of pain intensity based on similarity measures, using EMG, ECG and SC features, together with meta-information.




\section{Methods}

\subsection{BioVid Heat Pain Database}

\begin{figure}
	\centering
	\includegraphics[width=0.45\textwidth]{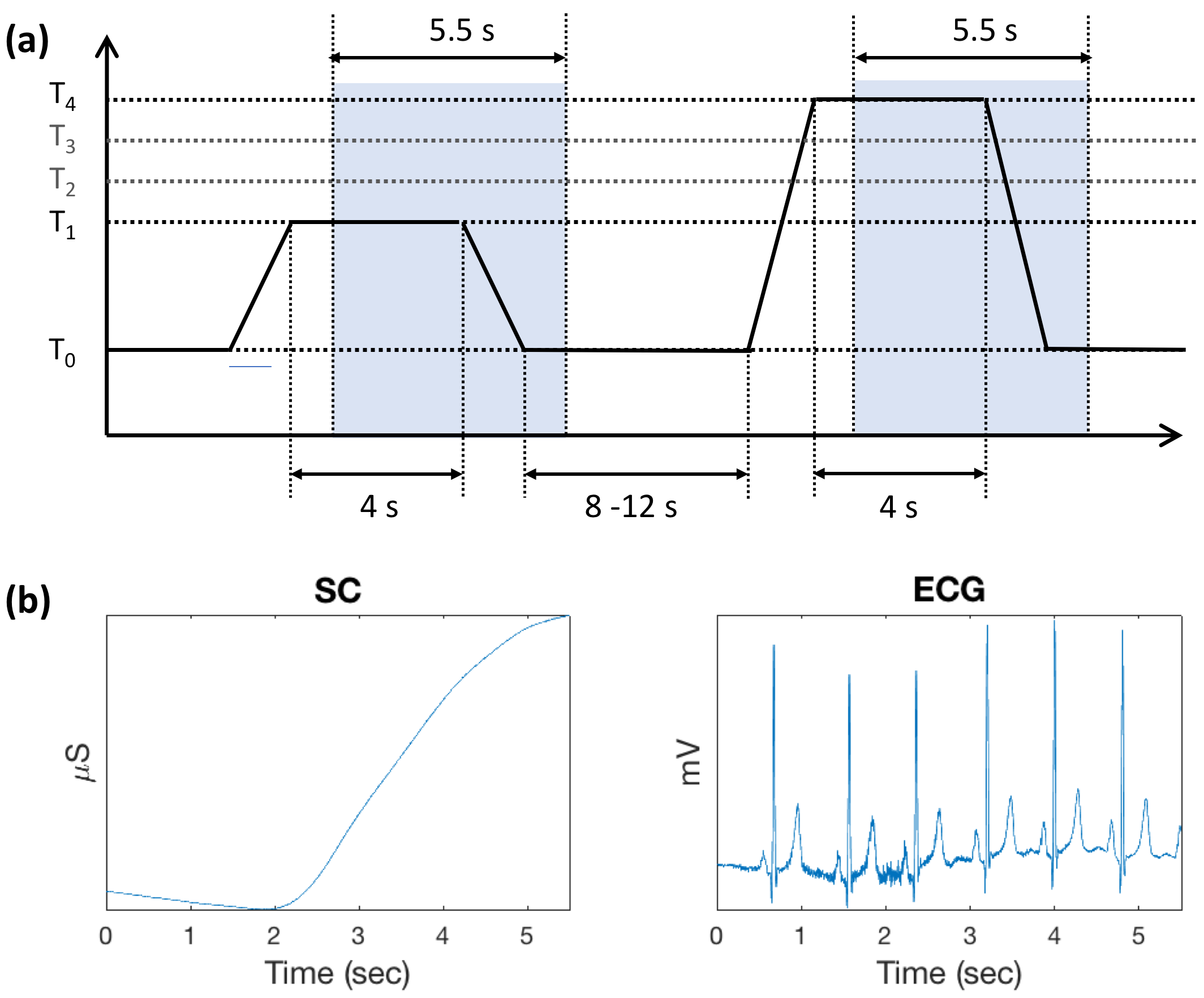}
    \caption{ \textbf{(a)} Pain stimulation, showing the heat levels ($T_0$ to $T_4$) and pause between stimuli. The features are extracted from the blue window of length 5.5 seconds. \textbf{(b)} Raw data streams from an exemplary pain stimulation with temperature $T_4$.}
    \label{fig:protocol}
\end{figure}

Our experiments were conducted with the BioVid Heat Pain database \cite{Walter2013}. The experimental setup consisted of a thermode that was used for  pain elicitation on the right arm.
Before the data recording was started, each subject's individual \textit{pain threshold} (the temperature for which the participant's sensing changes from heat to pain) and  \textit{tolerance threshold} (the temperature at which the pain becomes unacceptable) were determined. These thresholds were used as the temperatures for the lowest and highest pain levels ($T_1$ and $T_4$ respectively) together with two additional intermediate levels ($T_2$ and $T_3$), thus obtaining four pain levels. These temperatures were equally distributed in the range between $T_1$ and $T_4$, and never exceeded $50.5^\circ$C.
For each subject, pain stimulation was applied 20 times for each of the 4 calibrated stimulation intensities ($T_1$ to $T_4$). Each of the different stimuli was applied for 4 seconds followed by a recovery phase randomized between 8-12 seconds (see Figure \ref{fig:protocol}). Together with 20 baseline measurements ($T_0=32^\circ$C), a total number of 100 stimulations were applied for each subject in randomized order. The following biosignals were recorded: (1) skin conductance, (2) electrocardiogram, (3) electromyogram. However, only the first two biosignals were used in this work. The dataset was pre-processed to extract windows of length 5.5 seconds starting 1 second after the target temperature was reached for each stimulation (see Figure \ref{fig:protocol}). In total, the dataset contained 8700 samples of length 5.5 seconds from 87 subjects, equally distributed in 5 classes: no pain (baseline, \texttt{BLN}; corresponding to $T_0$), and pain levels \texttt{P1} to \texttt{P4} (corresponding to calibrated temperatures from $T_1$ to $T_4$ ). Therefore, the dataset contained 20 samples per class and subject.


\subsection{Feature extraction}

Noxious stimuli and the resulting pain affect the activity of the two branches of the autonomic nervous system: the sympathetic nervous system and the parasympathetic nervous system \cite{benaroch2001}. Changes in sympathetic and parasympathetic tone can be detected using computationally traceable measures of skin conductance and heart rate variability \cite{Cowen2015}.
Based on previous work \cite{Werner2014a}, we compute the following features of SC and ECG, with some novel additions. All features were computed on the pre-processed windows of 5.5 seconds, and were subsequently standarized.

\subsubsection{Skin conductance} This is a measure for the electrical conductance of the skin based on the activity of the perspiratory glands. The skin conductance (SC) signal was measured from two electrodes positioned on the index and ring fingers. The following features were extracted: (1) maximum; (2) range; (3) standard deviation; (4) inter-quartile range; (5) root mean square; (6) mean; (7) mean absolute value of the first differences, that is, $mavfd(x)=\frac{1}{N-1} \sum_{i=1}^{N-1}|x_{i+1}-x_i|$;  (8) mean absolute value of the second differences, that is, $mavfsd(x)=\frac{1}{N-2} \sum_{i=1}^{N-2}|x_{i+2}-x_i|$; (9) mean absolute value of the first differences of the standardized signal $x^* = \frac{x-mean(x)}{std(x)}$; (10) mean absolute value of the second differences of the standardized signal $x^*$; (11) skewness; (12) kurtosis.

\subsubsection{Electrocardiogram features}

The electrocardiogram (ECG), that is, the voltage generated by the heart muscle during heartbeats, was measured on the skin using two electrodes, one on the upper right and one on the lower left of the body. The ECG was first filtered using a Butterworth bandpass filter with frequency range [0.1, 250] Hz. This reduces noise and removes baseline drift. The R waves of the ECG were detected by an automatic algorithm based on Pan-Tompkins algorithm for QRS complex detection \cite{Pan1985}. The inter-beat interval (IBI) signal was constructed from consecutive heart beats. Then, the following ECG features were extracted from the signal: (1) the mean of the IBIs; (2) the root mean square of the successive differences (RMSSD); (3) the mean of the standard deviations of the IBIs (SDNN); (4) the slope of the liner regression of IBIs in its time series; (5) the ratio of SDNN to RMSSD.

\subsection{Multi-task learning with neural networks}

Our algorithm employs neural networks, a nonlinear classifier in which each layer $i$ in the network performs the transformation 
$
\textbf{x}_{i+1} = \sigma (\textbf{W}_i\textbf{x}_i + \textbf{b})
$,
where $\textbf{x}_i$ represents the input of the $i$-th layer of the network ($\textbf{x}_0$ is the feature vector), $\textbf{W}_i$ and $\textbf{b}_i$ are the weight matrix and bias, and $\sigma$ is the activation function (here, rectified linear unit). The last layer of the network $\textbf{x}_N$ is fed to a sigmoid function that predicts the binary label $y$ corresponding to the input $\textbf{x}_0$.

In single-task neural networks (ST-NN), a backpropagation algorithm is used to minimize a single loss function (here, binary cross-entropy). Multi-task learning (MTL), on the other hand, involves the simultaneous training of two or more related tasks over shared representations. Therefore, a multi-task neural network (MT-NN) contains $M$ sigmoid classifiers, one for each task, and the optimization of the corresponding loss functions is done simultaneously. 

\begin{figure}
	\centering
	\includegraphics[width=0.455\textwidth]{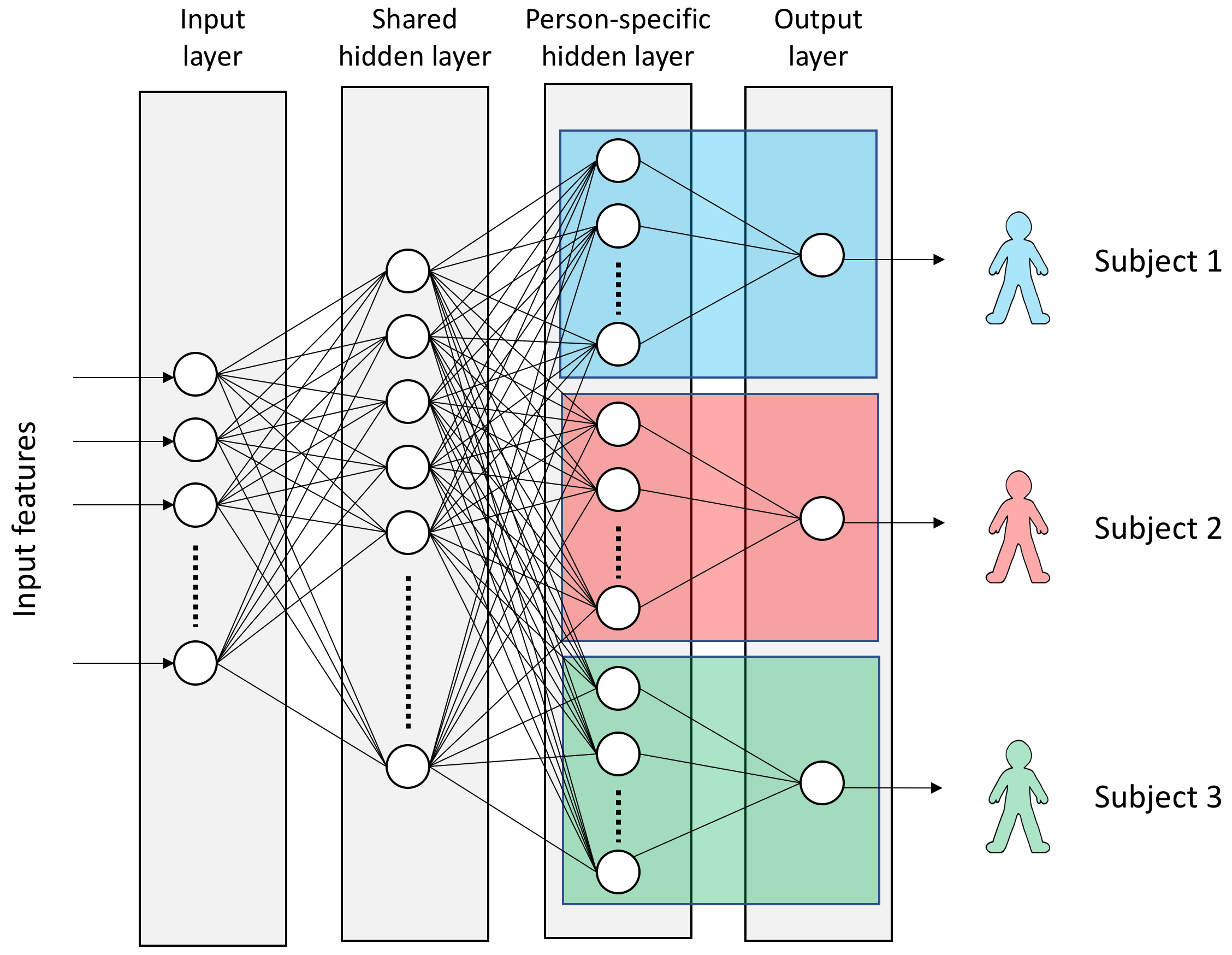}
    \caption{A simplified version of our multi-task neural network with two hidden layers: one shared, one person-specific.}
    \label{fig:nnarchitecture}
\end{figure}

In MT-NNs, MTL is typically done with either hard or soft parameter sharing of hidden layers \cite{Ruder2017}. In this work we use hard parameter sharing;
therefore, our MT-NN architecture shares some hidden layers between all tasks.  Like in Jaques et al.\cite{Jaques15}, the final layers are person-specific as can be seen in Figure \ref{fig:nnarchitecture}.
To regularize the network, we imposed an upper bound constraint on the norm of the network weights, applied dropout \cite{Srivastava2014}, and employed a validation-based early stopping strategy.
The whole algorithm was implemented using deep learning frameworks TensorFlow 1.2.1 and Keras 2.0.6.

\section{Results}

Our  experiments were conducted with the BioVid Heat Pain Database \cite{Walter2013}. We first tested two commonly used single-task classifiers: logistic regression (LR), and support vector machine with both linear kernel (SVM-L) and radial basis function kernel (SVM-RBF). SVMs have previously been used for pain recognition in \cite{Chu2017,Walter2014a,Werner2015a,Kachele2015}. The pain detection performance of these algorithms is summarized in Table \ref{tab:performance}, where classification accuracy is estimated via 10-fold cross-validation. 
The results indicate that SC features significantly outperform ECG features. 

The performances of both the ST-NN and MT-NN algorithms are also shown in Table \ref{tab:performance}. In both cases, we optimized the number of hidden layers and layer units. The best performance was achieved with a simple, fully-connected architecture  with two hidden layers. For the MT-NN, we defined one task for each subject in the dataset, and the network architecture consisted of one shared hidden layer and one task-specific hidden layer (see Figure \ref{fig:nnarchitecture}). For pain levels \texttt{P4} and \texttt{P3}, our MT-NN algorithm consistently outperformed any other algorithm, achieving a classification accuracy of 82.75\% and 70.04\% respectively (with both SC and ECG features). To put this in context, \cite{Werner2014a} reported an accuracy of 74.1\% and 65.0\% for \texttt{P4} and \texttt{P3} respectively. However, it must be noted that \cite{Werner2014a} also used EMG features, in addition to SC and ECG. In our work, we only used SC and ECG features.

\begin{table}[]
\centering
\caption{Pain detection with multi-modal data. Mean accuracies and standard deviations are reported for 10-fold cross-validation, for different algorithms and classification tasks (different pain levels).}
\label{tab:performance}
\begin{tabular}{|l|l|c|c|c|}
\hline
\multirow{2}{*}{\begin{tabular}[l]{@{}l@{}}Binary\\ classification\end{tabular}} & \multirow{2}{*}{  ML } & \multicolumn{3}{l|}{Used Features} \\
 &  & SC & ECG & SC+ECG \\ \hline \hline
\multirow{4}{*}{BLN vs P4} & LR & 77.67(1.44)  &  58.68(2.26) & 77.95(2.61)  \\
 & SVM-L &77.87(3.05)  & 58.82(2.27)  & 78.05(2.25) \\
 & SVM-RBF & 77.79(2.17)  &  58.25(2.20)  &  78.02(1.98) \\
 & ST-NN & 78.16(1.60) & 59.32(1.26) & 78.16(1.63) \\
 & MT-NN & \textbf{79.98(0.92)} & \textbf{62.50(1.69)} &  \textbf{82.75(1.86)}\\ \hline \hline
\multirow{4}{*}{BLN vs P3} & LR & 67.17(2.58)  & 54.77(2.69) & 67.39(2.13) \\
 & SVM-L & 67.38(2.64) & 55.00(2.07)  &  67.76(1.66) \\
 & SVM-RBF & 66.84(1.99) &  54.43(1.70)  &  67.50(3.34) \\
 & ST-NN & 67.93(0.82) & 55.67(2.52) & 67.75(0.62) \\
 & MT-NN & \textbf{69.76(1.66)} & \textbf{56.03(1.79)} &  \textbf{70.04(1.68)} \\ \hline \hline
\multirow{4}{*}{BLN vs P2} & LR & 58.79(3.36)  & 51.81(2.56) & 58.99(1.43)  \\
 & SVM-L & 58.56(2.99) & 51.75(2.88) &  59.51(2.41) \\
 & SVM-RBF & 58.22(3.26) & 51.90(2.34) &  57.87(2.68) \\
 & ST-NN &  59.74(1.11) & 51.87(1.66) & \textbf{59.71(1.50)}  \\
 & MT-NN & \textbf{60.34(1.30)} & \textbf{53.17(1.43)} &  58.62(0.61)\\ \hline \hline
\multirow{4}{*}{BLN vs P1} & LR &  53.13(2.81)& \textbf{50.69(2.30)}  & 53.30(2.51)  \\
 & SVM-L & \textbf{53.33(2.14)} & 50.55(2.08)  & \textbf{54.22(2.84)} \\
 & SVM-RBF & 52.53(1.79) & 48.65(3.56)   & 52.07(4.05)  \\
 & ST-NN & 52.33(1.41) & 47.35(0.30) & 52.53(2.42) \\
 & MT-NN & 50.01(1.81)  & 48.57(1.76) & 51.72(1.15) \\ \hline
\end{tabular}
\end{table}

\section{Discussion}
Most previous approaches to pain recognition are based on  subject-independent, or population, models: a one-fits-all pain recognition model that merges all the data from the available training population.
Unfortunately, population models often exhibit weak performance when applied to test data for new users due to inter-subject heterogeneity. In the context of pain, autonomic responses due to  pain have been shown to vary across subjects due to subject-specific differences (e.g. gender) \cite{Ledowski2011,Tousignant-laflamme2005,Tousignant-Laflamme2006}. Therefore, in this work we investigated the use of  multi-task learning to account for these subject-specific differences. Furthermore, we used skin conductance and heart rate features only, since these may be obtained from wrist-sensors, and therefore pain recognition methods based on these features may be used to develop a wearable device for pain monitoring that is suitable for daily use in clinical settings.

Our work has demonstrated that accounting for individual differences through MTL results in improved pain intensity recognition performance compared to other approaches (e.g. \cite{Werner2014a}), even though we only used SC and ECG features. Future work will explore how to better define the tasks (e.g. using similarity measures and meta-information \cite{Kachele2016}), extract better  features, and will also investigate the validity of this approach in real clinical settings.




%

{\small
\bibliographystyle{IEEEtran}
\bibliography{Mendeley}

\begin{thebibliography}{10}
\providecommand{\url}[1]{#1}
\csname url@samestyle\endcsname
\providecommand{\newblock}{\relax}
\providecommand{\bibinfo}[2]{#2}
\providecommand{\BIBentrySTDinterwordspacing}{\spaceskip=0pt\relax}
\providecommand{\BIBentryALTinterwordstretchfactor}{4}
\providecommand{\BIBentryALTinterwordspacing}{\spaceskip=\fontdimen2\font plus
\BIBentryALTinterwordstretchfactor\fontdimen3\font minus
  \fontdimen4\font\relax}
\providecommand{\BIBforeignlanguage}[2]{{%
\expandafter\ifx\csname l@#1\endcsname\relax
\typeout{** WARNING: IEEEtran.bst: No hyphenation pattern has been}%
\typeout{** loaded for the language `#1'. Using the pattern for}%
\typeout{** the default language instead.}%
\else
\language=\csname l@#1\endcsname
\fi
#2}}
\providecommand{\BIBdecl}{\relax}
\BIBdecl

\bibitem{DeCWilliams2016}
A.~C. d.~C. Williams and K.~D. Craig, ``{Updating the definition of pain},''
  \emph{PAIN}, vol. 157, no.~11, pp. 2420--2423, 11 2016.

\bibitem{Younger2009}
J.~Younger, R.~McCue, and S.~Mackey, ``{Pain outcomes: A brief review of
  instruments and techniques},'' \emph{Current Pain and Headache Reports},
  vol.~13, no.~1, pp. 39--43, 2009.

\bibitem{LopezMartinez2017c}
\BIBentryALTinterwordspacing
D.~Lopez~Martinez, O.~Rudovic, and R.~Picard, ``{Personalized Automatic
  Estimation of Self-reported Pain Intensity from Facial Expressions},'' in
  \emph{Computer Vision and Pattern Recognition}, Hawaii, 2017. [Online].
  Available: \url{http://arxiv.org/abs/1706.07154}
\BIBentrySTDinterwordspacing

\bibitem{Chu2017}
Y.~Chu, X.~Zhao, J.~Han, and Y.~Su, ``{Physiological Signal-Based Method for
  Measurement of Pain Intensity},'' \emph{Frontiers in Neuroscience}, vol.~11,
  no. May, pp. 1--13, 5 2017.

\bibitem{Kachele2016}
M.~Kachele, P.~Thiam, M.~Amirian, F.~Schwenker, and G.~Palm, ``{Methods for
  Person-Centered Continuous Pain Intensity Assessment From Bio-Physiological
  Channels},'' \emph{IEEE Journal of Selected Topics in Signal Processing},
  vol.~10, no.~5, pp. 854--864, 8 2016.

\bibitem{benaroch2001}
E.~E. Benarroch, ``{Pain-autonomic interactions: A selective review},''
  \emph{Clinical Autonomic Research}, vol.~11, no.~6, pp. 343--349, 12 2001.

\bibitem{Walter2014a}
S.~Walter, S.~Gruss, K.~Limbrecht-Ecklundt, H.~C. Traue, P.~Werner,
  A.~Al-Hamadi, N.~Diniz, G.~M. da~Silva, and A.~O. Andrade, ``{Automatic pain
  quantification using autonomic parameters},'' \emph{Psychology and
  Neuroscience}, vol.~7, no.~3, pp. 363--380, 2014.

\bibitem{Werner2015a}
S.~Gruss, R.~Treister, P.~Werner, H.~C. Traue, S.~Crawcour, A.~Andrade, and
  S.~Walter, ``{Pain Intensity Recognition Rates via Biopotential Feature
  Patterns with Support Vector Machines},'' \emph{PLOS ONE}, vol.~10, no.~10,
  10 2015.

\bibitem{Zhang2012}
Z.~G. Zhang, L.~Hu, Y.~S. Hung, A.~Mouraux, and G.~D. Iannetti, ``{Gamma-Band
  Oscillations in the Primary Somatosensory Cortex--A Direct and Obligatory
  Correlate of Subjective Pain Intensity},'' \emph{Journal of Neuroscience},
  vol.~32, no.~22, pp. 7429--7438, 5 2012.

\bibitem{Ledowski2011}
T.~Ledowski, J.~Stein, S.~Albus, and B.~MacDonald, ``{The influence of age and
  sex on the relationship between heart rate variability, haemodynamic
  variables and subjective measures of acute post-operative pain},''
  \emph{European Journal of Anaesthesiology}, vol.~28, no.~6, pp. 433--437, 6
  2011.

\bibitem{Tousignant-laflamme2005}
Y.~Tousignant-Laflamme, P.~Rainville, and S.~Marchand, ``{Establishing a Link
  Between Heart Rate and Pain in Healthy Subjects: A Gender Effect},''
  \emph{The Journal of Pain}, vol.~6, no.~6, pp. 341--347, 6 2005.

\bibitem{Tousignant-Laflamme2006}
Y.~Tousignant-Laflamme and S.~Marchand, ``{Sex differences in cardiac and
  autonomic response to clinical and experimental pain in LBP patients},''
  \emph{European Journal of Pain}, vol.~10, no.~7, pp. 603--614, 2006.

\bibitem{Jaques15}
N.~Jaques, S.~Taylor, E.~Nosakhare, A.~Sano, and R.~Picard, ``{Multi-task
  Learning for Predicting Health, Stress, and Happiness},'' in \emph{NIPS
  Workshop on Machine Learning for Healthcare}, 2016.

\bibitem{Ruder2017}
\BIBentryALTinterwordspacing
S.~Ruder, ``{An Overview of Multi-Task Learning in Deep Neural Networks},''
  2017. [Online]. Available: \url{http://arxiv.org/abs/1706.05098}
\BIBentrySTDinterwordspacing

\bibitem{Storm2008}
H.~Storm, ``{Changes in skin conductance as a tool to monitor nociceptive
  stimulation and pain},'' \emph{Current Opinion in Anaesthesiology}, vol.~21,
  no.~6, pp. 796--804, 12 2008.

\bibitem{Jess2016}
G.~Jess, E.~M. Pogatzki-Zahn, P.~K. Zahn, and C.~H. Meyer-Frieem, ``{Monitoring
  heart rate variability to assess experimentally induced pain using the
  analgesia nociception index},'' \emph{European Journal of Anaesthesiology},
  vol.~33, no.~2, pp. 118--125, 2016.

\bibitem{Sacco2013}
M.~Sacc{\`{o}}, M.~Meschi, G.~Regolisti, S.~Detrenis, L.~Bianchi,
  M.~Bertorelli, S.~Pioli, A.~Magnano, F.~Spagnoli, P.~G. Giuri, E.~Fiaccadori,
  and A.~Caiazza, ``{The Relationship Between Blood Pressure and Pain},''
  \emph{The Journal of Clinical Hypertension}, vol.~15, no.~8, pp. 600--605, 8
  2013.

\bibitem{Hamunen2012}
K.~Hamunen, V.~Kontinen, E.~Hakala, P.~Talke, M.~Paloheimo, and E.~Kalso,
  ``{Effect of pain on autonomic nervous system indices derived from
  photoplethysmography in healthy volunteers},'' \emph{British Journal of
  Anaesthesia}, vol. 108, no.~5, pp. 838--844, 5 2012.

\bibitem{Aasted2016}
C.~M. Aasted, M.~A. Yucel, S.~C. Steele, K.~Peng, D.~A. Boas, L.~Becerra, and
  D.~Borsook, ``{Frontal lobe hemodynamic responses to painful stimulation: A
  potential brain marker of nociception},'' \emph{PLoS ONE}, vol.~11, no.~11,
  pp. 1--12, 2016.

\bibitem{Lotsch2014}
J.~L{\"{o}}tsch, B.~G. Oertel, and A.~Ultsch, ``{Human models of pain for the
  prediction of clinical analgesia},'' \emph{Pain}, vol. 155, no.~10, pp.
  2014--2021, 2014.

\bibitem{Treister2012}
R.~Treister, M.~Kliger, G.~Zuckerman, I.~G. Aryeh, and E.~Eisenberg,
  ``{Differentiating between heat pain intensities: The combined effect of
  multiple autonomic parameters},'' \emph{Pain}, vol. 153, no.~9, pp.
  1807--1814, 2012.

\bibitem{Chu2014}
Y.~Chu, X.~Zhao, J.~Yao, Y.~Zhao, and Z.~Wu, ``{Physiological Signals Based
  Quantitative Evaluation Method of the Pain},'' \emph{IFAC Proceedings
  Volumes}, vol.~47, no.~3, pp. 2981--2986, 2014.

\bibitem{Walter2013}
S.~Walter, S.~Gruss, H.~Ehleiter, J.~Tan, H.~C. Traue, S.~Crawcour, P.~Werner,
  A.~Al-Hamadi, and A.~O. Andrade, ``{The biovid heat pain database data for
  the advancement and systematic validation of an automated pain recognition
  system},'' in \emph{2013 IEEE International Conference on Cybernetics
  (CYBCO)}.\hskip 1em plus 0.5em minus 0.4em\relax IEEE, 6 2013, pp. 128--131.

\bibitem{Werner2014a}
P.~Werner, A.~Al-Hamadi, R.~Niese, S.~Walter, S.~Gruss, and H.~C. Traue,
  ``{Automatic Pain Recognition from Video and Biomedical Signals},'' in
  \emph{2014 22nd International Conference on Pattern Recognition}.\hskip 1em
  plus 0.5em minus 0.4em\relax IEEE, 8 2014, pp. 4582--4587.

\bibitem{Kachele2015}
M.~K{\"{a}}chele, P.~Werner, A.~Al-Hamadi, G.~Palm, S.~Walter, and
  F.~Schwenker, ``{Bio-Visual Fusion for Person-Independent Recognition of Pain
  Intensity},'' in \emph{Multiple Classifier Systems}, S.~F., R.~F., and K.~J.,
  Eds.\hskip 1em plus 0.5em minus 0.4em\relax Springer, 2015, vol. 9132, pp.
  220--230.

\bibitem{Iliadis2015}
M.~K{\"{a}}chele, P.~Thiam, M.~Amirian, P.~Werner, S.~Walter, F.~Schwenker, and
  G.~Palm, ``{Multimodal Data Fusion for Person-Independent, Continuous
  Estimation of Pain Intensity},'' in \emph{Communications in Computer and
  Information Science}, L.~Iliadis and C.~Jayne, Eds.\hskip 1em plus 0.5em
  minus 0.4em\relax Cham: Springer, 2015, vol. 517, pp. 275--285.

\bibitem{Cowen2015}
R.~Cowen, M.~K. Stasiowska, H.~Laycock, and C.~Bantel, ``{Assessing pain
  objectively: the use of physiological markers},'' \emph{Anaesthesia},
  vol.~70, no.~7, pp. 828--847, 7 2015.

\bibitem{Pan1985}
J.~Pan and W.~J. Tompkins, ``{A Real-Time QRS Detection Algorithm},''
  \emph{IEEE Transactions on Biomedical Engineering}, vol. BME-32, no.~3, pp.
  230--236, 3 1985.

\bibitem{Srivastava2014}
N.~Srivastava, G.~Hinton, A.~Krizhevsky, I.~Sutskever, and R.~Salakhutdinov,
  ``{Dropout : A Simple Way to Prevent Neural Networks from Overfitting},''
  \emph{Journal of Machine Learning Research}, vol.~15, pp. 1929--1958, 2014.

\end{thebibliography}
}

\end{document}